\documentclass[twocolumn,aps,noshowkeys,noshowpacs,superscriptaddress]{revtex4}

	\usepackage{natbib}
	\usepackage{amsmath}
	\usepackage{makeidx}
	\usepackage{amsfonts}
	\usepackage[ansinew]{inputenc}
	\usepackage[usenames,dvipsnames]{pstricks}
	\usepackage{subfigure}
	\usepackage{epsfig}
	\usepackage{pst-grad} 
	\usepackage{pst-plot} 
	\usepackage{epstopdf}
	\usepackage{multirow}
	\usepackage{titlesec}
	\usepackage{float}
	\usepackage{amssymb}
	\usepackage{siunitx}

	\titlespacing{\section}{0pt}{5pt}{5pt}
	\titlespacing{\subsection}{0pt}{5pt}{5pt}
	\titlespacing{\subsubsection}{0pt}{5pt}{5pt}

\makeindex

\begin{document}

\title{The photon pair source that survived a rocket explosion}

\author{Zhongkan Tang}
\affiliation{Centre for Quantum Technologies, National University of Singapore,\\ Block S15, 3 Science Drive 2, 117543 Singapore.}
\author{Rakhitha Chandrasekara}
\affiliation{Centre for Quantum Technologies, National University of Singapore,\\ Block S15, 3 Science Drive 2, 117543 Singapore.}
\author{Yue Chuan Tan}
\affiliation{Centre for Quantum Technologies, National University of Singapore,\\ Block S15, 3 Science Drive 2, 117543 Singapore.}
\author{Cliff Cheng}
\affiliation{Centre for Quantum Technologies, National University of Singapore,\\ Block S15, 3 Science Drive 2, 117543 Singapore.}
\author{Kadir Durak}
\affiliation{Centre for Quantum Technologies, National University of Singapore,\\ Block S15, 3 Science Drive 2, 117543 Singapore.}
\author{Alexander Ling}
\affiliation{Centre for Quantum Technologies, National University of Singapore,\\ Block S15, 3 Science Drive 2, 117543 Singapore.}
\affiliation{Department of Physics, National University of Singapore,\\ 3 Science Drive 2,  117551 Singapore}


\begin{abstract}
We report on the performance of a compact photon pair source that was recovered intact from a failed space launch. The source had been embedded in a nanosatellite and was designed to perform pathfinder experiments leading to  global quantum communication networks using spacecraft. Despite the launch vehicle explosion soon after takeoff, the nanosatellite was successfully retrieved from the accident site and the source within it was found to be fully operational. We describe the assembly technique for the rugged source. Post-recovery data is compared to baseline measurements collected before the launch attempt and no degradation in brightness or polarization correlation was observed. The survival of the source through an extreme environment provides strong evidence that it is possible to engineer rugged quantum optical systems.

\end{abstract}

\keywords{Nonlinear optics, parametric processes, quantum optics, optical design of instruments}

\maketitle

Secure generation of symmetric key material at distant sites using quantum signals, known as quantum key distribution (QKD), is one of the most technologically mature outcomes of research into quantum communication. In particular, entanglement-based QKD~\cite{Ekert1991} is a powerful technique that relies on the quantum entanglement between photons.

Current QKD networks have a distance limit due to fiber losses~\cite{Gisin2002, Edo2002} and the lack of quantum repeaters~\cite{Briegel1998}. To overcome the distance limit and to build regional and global QKD networks, several efforts are ongoing to utilize satellites as receivers or transmitters~\cite{Jennewein2014, Bourgoin2013, Wang2013, Merali2012, Yin2013}. We have proposed that a cost-effective approach to this challenge is to capitalize on emerging nanosatellite technology such as CubeSats~\cite{Woellert2011} in order to perform technology validation experiments in small iterative steps~\cite{Morong2012}. 

One of the driving requirements in this approach is to develop compact and rugged sources of entangled photon pairs that are compatible with the size, weight and power requirements of small satellites. This is a challenge because the workhorse method for entangled photon pairs requires the use of precisely aligned bulk optics relying on spontaneous parametric downconversion (SPDC)~\cite{Burnham1970}. We identified a promising source design~\cite{Trojek2008} that was compact (due to collinear emission) and did not require stringent temperature control, thus reducing weight and power requirements. Besides the source, opto-electronics components such as single photon detectors and polarization analyzers also had to be validated in a space environment.

\begin{figure}[t]
    \centering
    \includegraphics[width=0.95\linewidth]{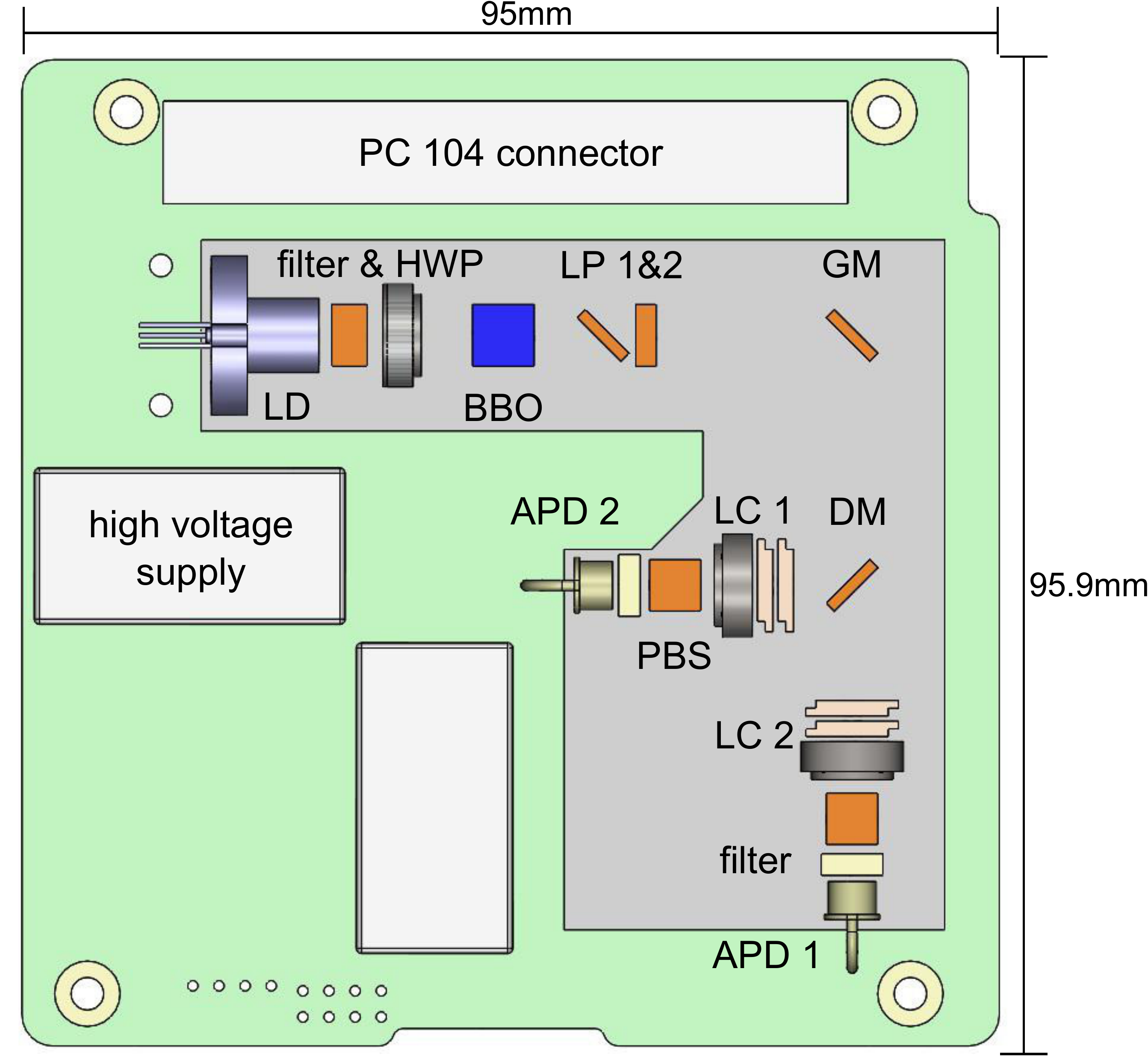}
    \caption{\label{fig:pack} Layout of the optical components in the science package. The optical unit (grey outline) is fixed to a printed circuit board (green outline) that provides the interface to the CubeSat. Collinear photon pairs at 760nm (signal) and 867nm (idler) are produced in a single $\beta$-barium-borate (BBO) crystal pumped with 405nm photons from an ONDAX laser diode (LD). The pump spectrum and polarization mode is prepared by band-pass filter and half-wave plate (HWP). Excess pump light is removed via two long-pass filters (LP 1 \& 2). The optical path is folded by a gold mirror (GM) and the non-degenerate photon pairs are split via a dichroic mirror (DM). Polarization analysis is achieved with liquid crystal polarization rotators (LC 1 \& 2) and polarizing beam splitters (PBS). Silicon avalanche photodiodes (APDs) are used for the detection of the pairs and each APD has an individual spectral filter and a high voltage supply.}
  \end{figure}

As a first step towards these validation experiments a science package that comprised of a correlated photon pair source (utilizing only one SPDC crystal) and all the necessary electronics was assembled. The photon correlations were sufficient to test the assembly technique as well as to observe how the single photon manipulation and detection apparatus could perform in space. The complete package was compact (\SI{9.5}{cm}$\times$\SI{9.6}{cm}$\times$\SI{3.8}{cm}) with a mass of about \SI{250}{g} and was designed to fit a standard CubeSat spacecraft. The optical layout is shown in Fig.~\ref{fig:pack}, while a complete description of the electronics is described in~\cite{Cheng2015}.

The science package was installed on the GomX-2 CubeSat which was to be deployed from the International Space Station~\cite{gomx2}. Unfortunately the mission failed when the launch vehicle was destroyed shortly after launch~\cite{NASA2015} but GomX-2 was successfully recovered from the debris. The science package within was found to be completely operational, continuing to produce high quality polarization correlations. In the rest of this paper we will describe the assembly technique for constructing the photon pair source. The performance of the source before and after recovery is also presented.

The photon pair source is built around type-I collinear SPDC. The pump laser is a grating stabilized \SI{405}{nm} laser diode. When in operation the pump power is stabilized and used to produce non-degenerate photon pairs at \SI{760}{nm} (signal) and \SI{867}{nm} (idler) in a \SI{6}{mm} long $\beta$-barium-borate (BBO) crystal. A dichroic mirror separates the signal and idler before each photon is interrogated by a polarization analyzer composed of a liquid crystal (LC) retarder and a polarizing filter. Photons transmitted by the filter are detected by passively quenched avalanche photodiodes (APDs). The average power consumed by the science package is approximately \SI{1.5}{W}. 

In the science package the optical and opto-electronic components are contained within a machined monolithic optical unit (black anodized Al6061). Each component is provided with a recessed pocket within the unit and is held in the pocket with epoxy (UHU\textsuperscript{\textregistered} 300). The SPDC photons were directly incident on the APDs without the use of collection lenses. The components most sensitive to alignment are the BBO crystal, and the two mirrors used to direct the photons to the detectors. In particular the BBO crystal had an angular tolerance at the order of $\pm$\SI{100}{\micro rad}. 

\begin{figure}[t]
	\begin{center}
	\includegraphics[scale=0.28]{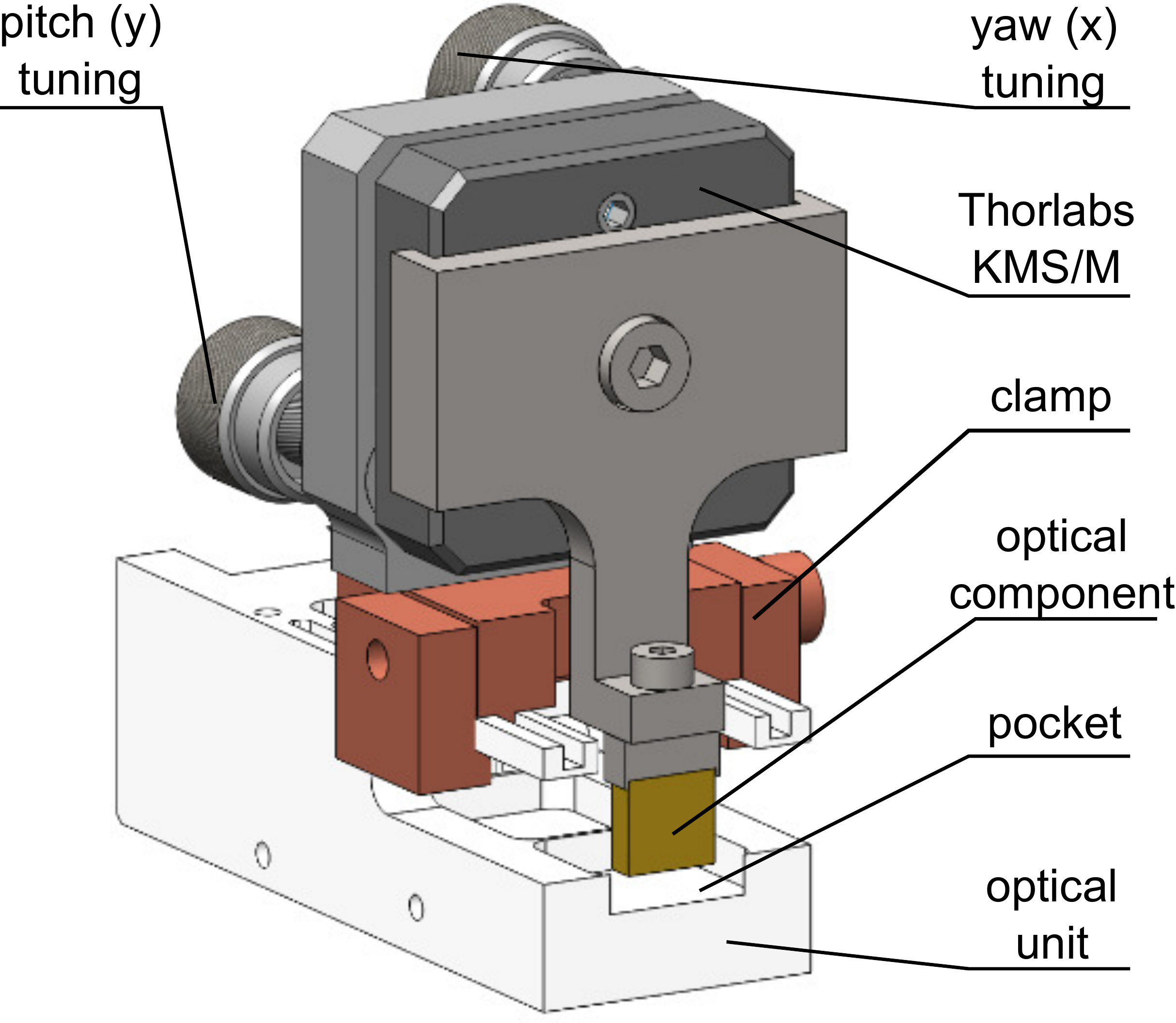}
	\caption{Toolkit used to align component in the optical unit. The yaw and pitch angle can be adjusted independently using the two knobs. The component is aligned within a pocket that is machined into the optical unit. Epoxy is applied to secure the component within this pocket after alignment. After the epoxy has cured, the adjuster is removed enabling the compact form-factor.}
	\label{fig:tilt}
	\end{center}
	\end{figure}

To align the components with the required accuracy a toolkit was developed around a standard optical adjuster (see Fig.~\ref{fig:tilt}).  The optical unit was designed with rails enabling the adjuster to slide into the correct position before being secured by a clamp. A custom holder attached the component to the adjuster (Thorlabs KMS/M) that provided fine adjustment in the x (or yaw) and y (or pitch) axes. This adjustment can be performed without contact between the component and unit. Once the alignment was finalized epoxy was injected into the pocket to secure the component. After the epoxy had cured the adjuster was released and removed. This method has also been used to build a polarization-entangled source. 

To maximize data collection, the science package was programmed with 16 settings enabling different types of thirty-minute-duration experiments. Each setting is associated with a unique command code. In the mission operations, ground controllers would uplink the desired experimental setting and the appropriate command code is then passed to the science package at the appropriate time. The GomX-2 mission could guarantee a minimum uninterrupted experimental run of thirty minutes every 24 hours. 

The primary experimental setting is used to observe the quality and brightness of the source. In this profile, the LC rotator for the idler photon (LC 2) is adjusted to maximize transmission through the polarizing filter. The LC rotator for the signal photon (LC 1) is then stepped through a series of voltage settings leading to a variation in the overall rate of coincident counts between the two detectors. The achieved contrast (or visibility) quantifies the quality of the correlations between the photon pairs.

The package has a secondary experimental setting that performs the same function, but swaps the function of LC 1 and LC 2. It also operates the pump laser with constant current instead of stabilizing the optical power. This provides a contingency scenario in case the PID-based power control in the primary setting becomes unuseable. The pump power in this setting is \SI{6}{mW} (50\% of the power in the primary setting). This secondary setting is also a useful diagnostic tool to evaluate if the polarization analyzers have been damaged. The package also contains settings that are used to track instrumentation performance, e.g. APD dark count rate or pump laser threshold current. Had the GomX-2 spacecraft been commissioned successfully, these profiles would have enabled in-orbit data collection on component performance over time.

Identical devices had been tested successfully in a near-space environment previously~\cite{Tang2014}. During the launch campaign, a more formal series of tests were conducted after the science package had been integrated into the GomX-2 spacecraft. The sequential series of tests were a mechanical vibration test followed by thermal cycling and concluding with a vacuum test.

The first mechanical vibration test was to determine that the integrated spacecraft did not have a resonant frequency between \SI{5}{Hz} to \SI{100}{Hz}. This step is usually known as a ``sine-sweep" where a simple sinusoidal vibration at fixed frequency is provided to the spacecraft at low amplitude. The second step was to determine that it could survive a vibration profile of 14-grms with frequency components between \SI{20}{Hz} and \SI{2000}{Hz} along all three axes of the spacecraft. The spacecraft was also tested in a thermal chamber where the temperature was cycled between -\SI{10}{\celsius} and \SI{40}{\celsius} every 100 minutes. This thermal cycling was continuous for 24 hours. The spacecraft was also subjected to a low pressure of approximately $10^{-5}$~mbar for 24 hour. 
The science package was not damaged by the qualification tests and the polarization correlation visibility was determined to be approximately $96 \pm 2 \%$ both before and after testing.

After the GomX-2 spacecraft was recovered the science package was activated and data was collected from the primary and secondary experimental settings. From a detailed analysis of the data it was possible to conclude that the photon pair source survived the launch vehicle explosion intact. A checklist is presented in Table.~\ref{prepost} which compares the performance of key components in the science package before and after the explosion. 

\begin{table}[t]
\centering
\begin{tabular}{|c|c|c}
\hline
Parameter            &  \multicolumn{1}{c|}{baseline} & \multicolumn{1}{c|}{post-recovery} \\ \hline
Pump threshold &  \multicolumn{1}{c|}{27.8mA}           & \multicolumn{1}{c|}{27.8mA}            \\ \hline
Stabilised pump power &  \multicolumn{1}{c|}{12mW}           & \multicolumn{1}{c|}{12mW}            \\ \hline
APD dark count rate &  \multicolumn{1}{c|}{13000 cps}   & \multicolumn{1}{c|}{14500 cps}    \\ \hline
\end{tabular}
\caption{Three important device parameters for the science package are the pump laser threshold current, the stabilized pump power and the APD darkcount rate. The parameter values before and after explosion are presented.}
\label{prepost}
\end{table}

In terms of source brightness a critical system parameter is the power of the pump laser diode. The target power in the primary experimental setting is \SI{12}{mW} and this was achieved in the recovered science package. Another important parameter is the threshold current for the pump laser and this was observed to be \SI{27.8}{mA} both before and after the explosion. The detection APDs in the recovered package exhibited a slight increase in the dark count rate in all the experimental runs, but this is attributed to the storage environment leading to an elevated internal temperature of the nanosatellite. The thermistors onboard the science package reported values that were consistent with other thermal sensors within the nanosatellite. Both types of memory (flash and EEPROM) modules used on the science package are also intact. 

\begin{figure}[t]
\centering
\includegraphics[width=0.99\linewidth]{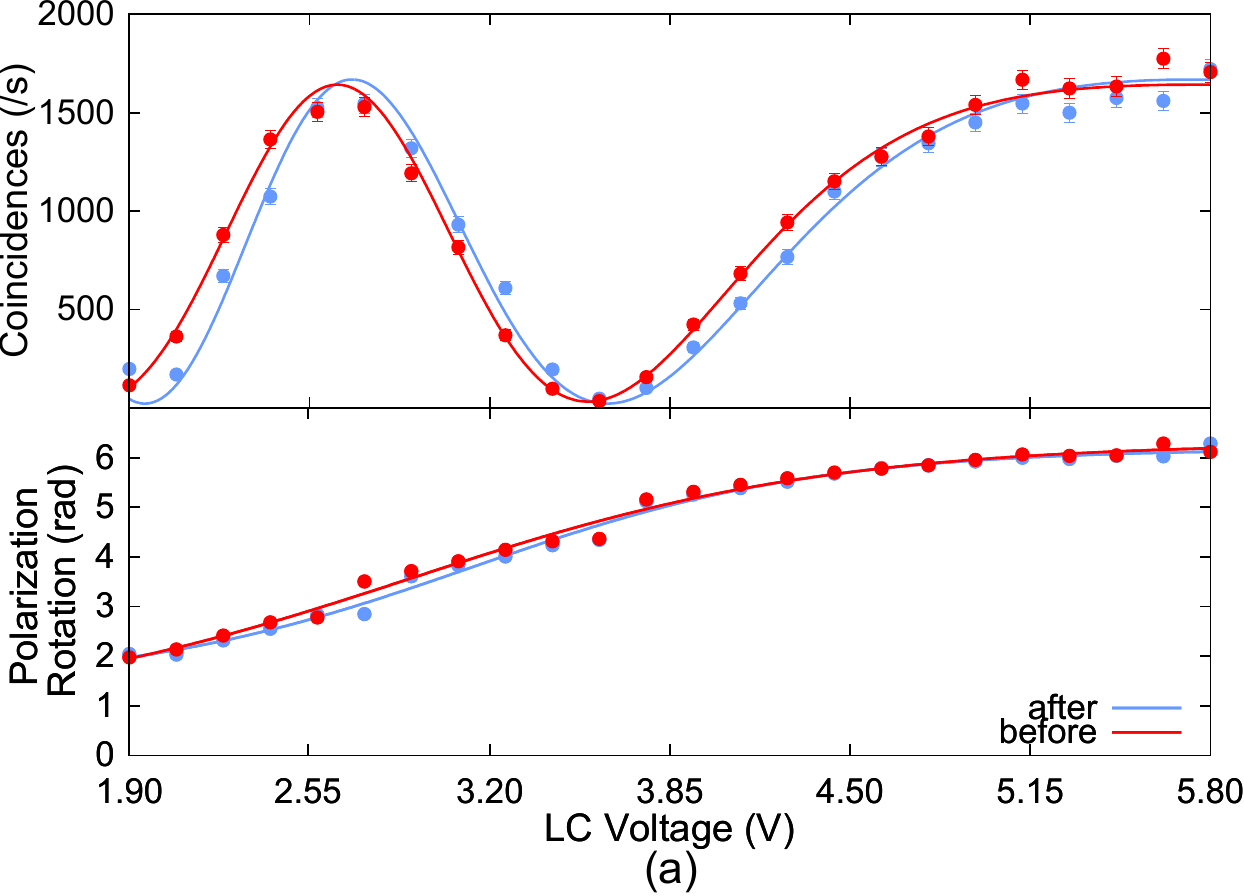} 
\hfill
\includegraphics[width=0.99\linewidth]{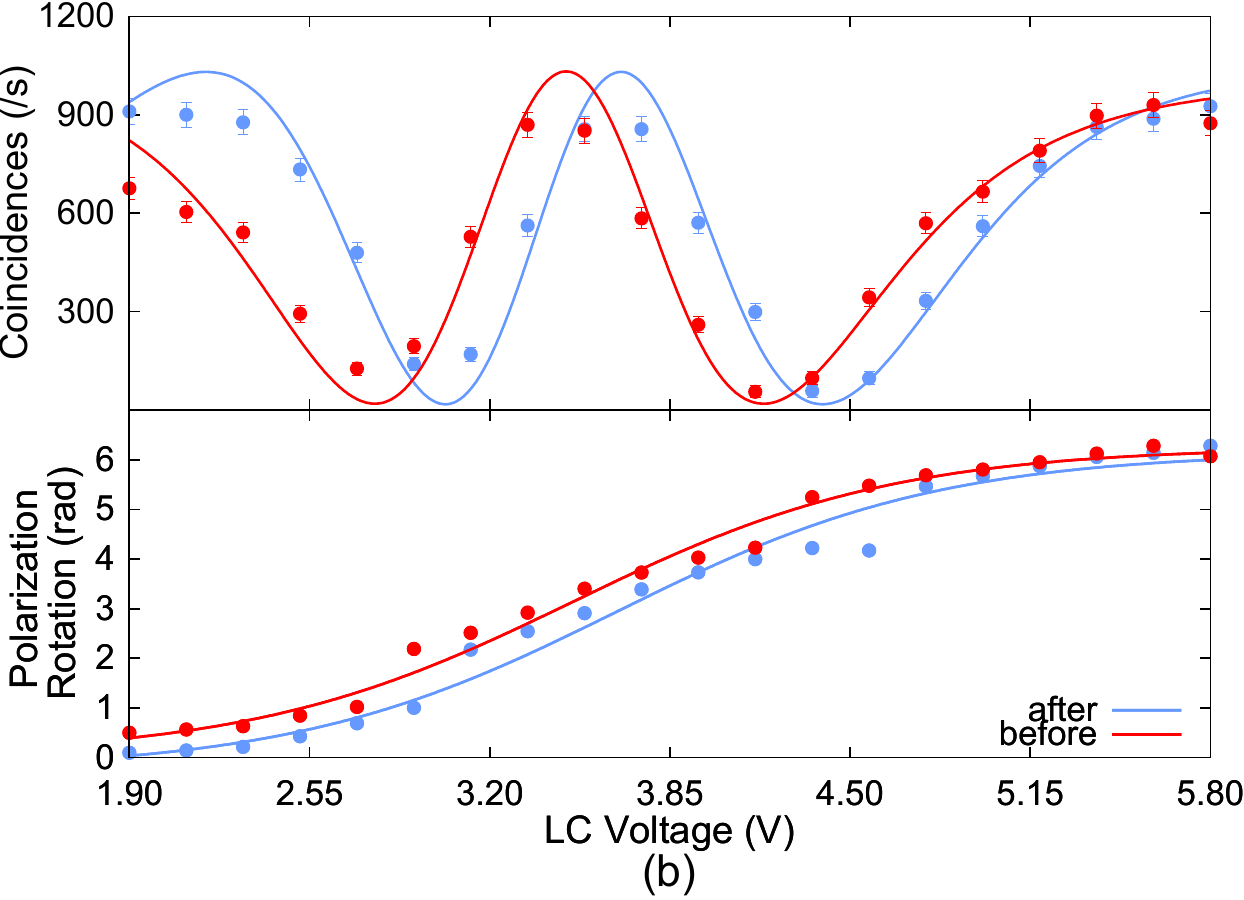}
\caption{Coincidence rate and calibration curve for the liquid crystal (LC) polarization analyzer for the primary (a) and secondary (b) experimental profiles. Both liquid crystals show a reduced ability to perform polarization rotation.}
\label{phaseshift}
\end{figure}

The photon correlations from the source before and after recovery for the primary and secondary experiment settings are shown in  Fig.~\ref{phaseshift} (a) and (b) respectively.
The figures also show the polarization rotation achieved for a given voltage setting to the liquid crystal devices.
Care was taken to compare data sets which were collected at the same temperature range. 
The brightness of the photon pair source is essentially unchanged while there is a notable shift in the position of the maxima and minima, especially for the experiments using the secondary settings.

These observed effects cannot be attributed to a misalignment of the photon pair source, but can be attributed instead to a degradation of components downstream of the SPDC process. This is because bandpass filters are present between the polarization rotators and the single photon detectors. Significant source misalignment would have resulted in collinear photons of a different wavelength. In such a scenario, the brightness would have been significantly reduced, while the position of the maxima and minima would have remained unchanged.

A plausible explanation for the observed phase shift is that the liquid crystal polarization rotators have suffered some damage. The shift in the maxima and minima positions correspond to a reduced ability of the liquid crystal devices to cause a polarization rotation for a given voltage setting.  Such a degradation in performance can be induced by mechanical or thermal shock. Another credible explanation is that the two independent electronic drivers for the liquid crystal devices have suffered some damage, but this is considered unlikely as there is no evidence of damage to other electronic sub-systems.

While no telemetry from the payload compartment during the explosion was available to the authors, it is possible to infer an approximate upper bound to the temperature experienced by the GomX-2 satellite. This is done by noting that the antenna deployment mechanism for the GomX-2 satellite was dependent on a cable-tie that had to melt at approximately \SI{144}{\celsius}. The satellite was recovered with the cable-tie intact making it very unlikely that the spacecraft internal components had reached this temperature. However, any temperature above \SI{90}{\celsius} would already have led to permanent change in the liquid crystals, so thermal-induced degradation cannot be ruled out. It is sometimes possible to restore the performance of the liquid crystal devices via a controlled process involving thermal and voltage treatment, but this requires access to the science package. A more conclusive statement on the cause of the liquid crystal performance may have to wait until the future disassembly of the GomX-2 satellite.

\begin{figure}[t]
    \centering
    \includegraphics[width=1\linewidth]{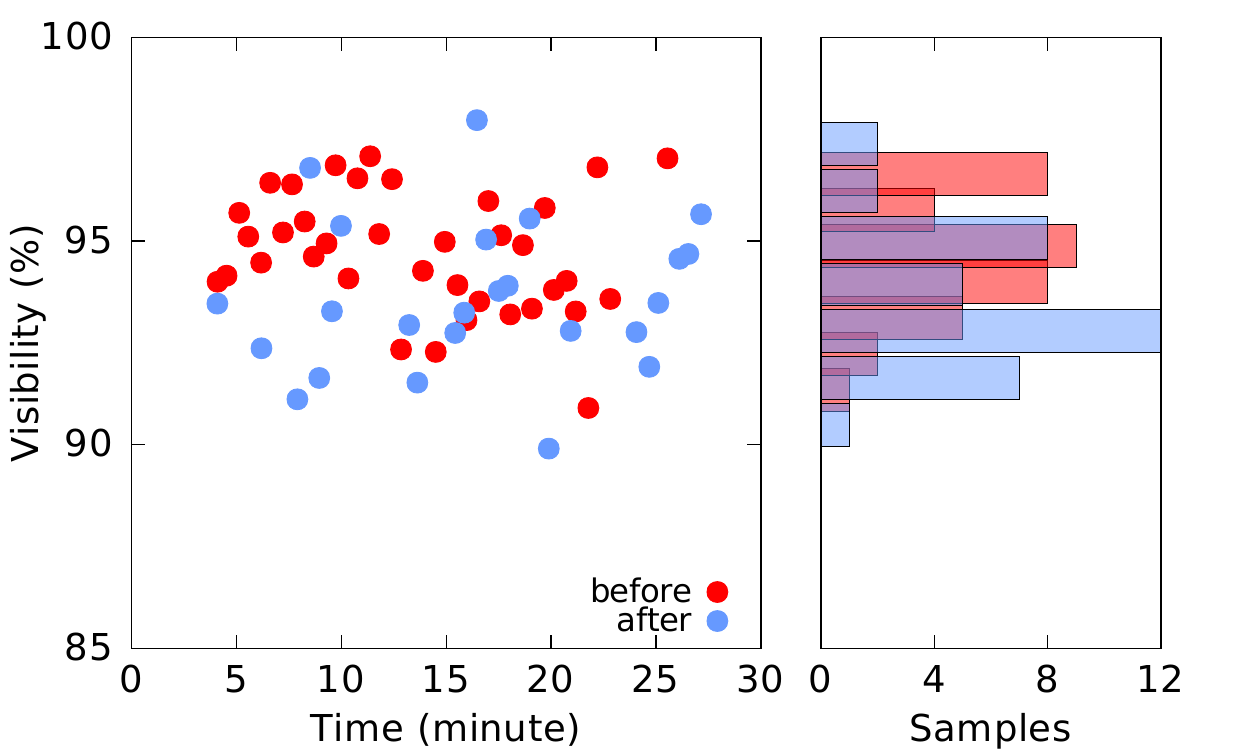}
    \caption{Comparison of polarization visibility before (red) and after (blue) explosion for a number of experimental runs taken over 30 minutes. The average baseline visibility is $95 \pm 1\%$ while the recovered package exhibited a visibility of $94 \pm 2\%$, showing no significant deviation after the explosion. The histogram on the right shows the distribution of the visibility.}
    \label{fig:comparehist}
\end{figure}

Despite the damage to the liquid crystal rotators, the science package was still able to demonstrate high polarization correlation between photon pairs. A summary of the visibility obtained from a number of experimental trials conducted using the recovered pair source is illustrated in Fig.~\ref{fig:comparehist} and compared to the visibility obtained in the baseline experiments. There is no drift in the visibility despite the change in liquid crystal performance.

We have presented the assembly technique and space qualification tests for a compact photon pair source that was embedded inside a nanosatellite that was subjected to a launch vehicle explosion. The recovery of the nanosatellite provided a unique opportunity to study how the photon pair source performed after undergoing such a very dramatic event.
The photon source and its accompanying electronic package was found to be operational. Despite some damage to the liquid crystal polarization rotators, the polarization correlation quality was essentially the same as baseline values collected before the launch attempt.
This data strongly suggests that quantum devices that are usually thought as sensitive instruments confined only to laboratory environments can actually be engineered to withstand very catastrophic environmental conditions.
Work is currently ongoing to prepare new and improved photon pair systems that exhibit improved ruggedness for deployment on future nanosatellite missions.

\begin{acknowledgements}
This work is supported by the National Research Foundation grant NRF-CRP12-2013-02 \emph{``Space based Quantum Key Distribution''}. The authors thank Yau Yong Sean for assistance on mechanical design on the optical unit. We acknowledge the support of the GomSpace ApS team in the post-recovery examination of the science package.
\end{acknowledgements}

\bibliographystyle{naturemag}
\bibliography{survival}

\begin{thebibliography}{10}
\expandafter\ifx\csname url\endcsname\relax
  \def\url#1{\texttt{#1}}\fi
\expandafter\ifx\csname urlprefix\endcsname\relax\def\urlprefix{URL }\fi
\providecommand{\bibinfo}[2]{#2}
\providecommand{\eprint}[2][]{\url{#2}}

\bibitem{Ekert1991}
\bibinfo{author}{Ekert, A.~K.}
\newblock \bibinfo{title}{{Quantum Cryptography Based on Bell's Theorem}}.
\newblock \emph{\bibinfo{journal}{Phys. Rev. Lett.}}
  \textbf{\bibinfo{volume}{67}}, \bibinfo{pages}{661--663}
  (\bibinfo{year}{1991}).

\bibitem{Gisin2002}
\bibinfo{author}{Gisin, N.}, \bibinfo{author}{Ribordy, G.},
  \bibinfo{author}{Tittel, W.} \& \bibinfo{author}{Zbinden, H.}
\newblock \bibinfo{title}{{Quantum cryptography}}.
\newblock \emph{\bibinfo{journal}{Rev. Mod. Phys.}}
  \textbf{\bibinfo{volume}{74}}, \bibinfo{pages}{145--195}
  (\bibinfo{year}{2002}).

\bibitem{Edo2002}
\bibinfo{author}{Waks, E.}, \bibinfo{author}{Zeevi, A.} \&
  \bibinfo{author}{Yamamoto, Y.}
\newblock \bibinfo{title}{{Security of Quantum Key Distribution with Entangled
  Photons Against Individual Attacks}}.
\newblock \emph{\bibinfo{journal}{Phys. Rev. A}} \textbf{\bibinfo{volume}{65}}
  (\bibinfo{year}{2002}).

\bibitem{Briegel1998}
\bibinfo{author}{Briegel, H.-J.}, \bibinfo{author}{D{\"{u}}r, W.},
  \bibinfo{author}{Cirac, J.} \& \bibinfo{author}{Zoller, P.}
\newblock \bibinfo{title}{{Quantum Repeaters: The Role of Imperfect Local
  Operations in Quantum Communication}}.
\newblock \emph{\bibinfo{journal}{Phys. Rev. Lett.}}
  \textbf{\bibinfo{volume}{81}}, \bibinfo{pages}{5932--5935}
  (\bibinfo{year}{1998}).

\bibitem{Jennewein2014}
\bibinfo{author}{Jennewein, T.} \emph{et~al.}
\newblock \bibinfo{title}{{The NanoQEY mission: ground to space quantum key and
  entanglement distribution using a nanosatellite}}.
\newblock \emph{\bibinfo{journal}{Proc. SPIE}} \textbf{\bibinfo{volume}{9254}},
  \bibinfo{pages}{925402} (\bibinfo{year}{2014}).

\bibitem{Bourgoin2013}
\bibinfo{author}{Bourgoin, J.-P.} \emph{et~al.}
\newblock \bibinfo{title}{{A comprehensive design and performance analysis of
  low Earth orbit satellite quantum communication}}.
\newblock \emph{\bibinfo{journal}{New J. Phys.}} \textbf{\bibinfo{volume}{15}},
  \bibinfo{pages}{023006} (\bibinfo{year}{2013}).

\bibitem{Wang2013}
\bibinfo{author}{Wang, J.-Y.} \emph{et~al.}
\newblock \bibinfo{title}{{Direct and full-scale experimental verifications
  towards ground-satellite quantum key distribution}}.
\newblock \emph{\bibinfo{journal}{Nature Photon.}}
  \textbf{\bibinfo{volume}{7}}, \bibinfo{pages}{387--393}
  (\bibinfo{year}{2013}).

\bibitem{Merali2012}
\bibinfo{author}{Merali, Z.}
\newblock \bibinfo{title}{{The quantum space race}}.
\newblock \emph{\bibinfo{journal}{Nature}} \textbf{\bibinfo{volume}{492}},
  \bibinfo{pages}{22--25} (\bibinfo{year}{2012}).

\bibitem{Yin2013}
\bibinfo{author}{Yin, J.} \emph{et~al.}
\newblock \bibinfo{title}{{Experimental quasi-single-photon transmission from
  satellite to earth}}.
\newblock \emph{\bibinfo{journal}{Opt. Express}} \textbf{\bibinfo{volume}{21}},
  \bibinfo{pages}{20032} (\bibinfo{year}{2013}).

\bibitem{Woellert2011}
\bibinfo{author}{Woellert, K.}, \bibinfo{author}{Ehrenfreund, P.},
  \bibinfo{author}{Ricco, A.~J.} \& \bibinfo{author}{Hertzfeld, H.}
\newblock \bibinfo{title}{{Cubesats: Cost-effective science and technology
  platforms for emerging and developing nations}}.
\newblock \emph{\bibinfo{journal}{Adv. Space Res.}}
  \textbf{\bibinfo{volume}{47}}, \bibinfo{pages}{663--684}
  (\bibinfo{year}{2011}).

\bibitem{Morong2012}
\bibinfo{author}{Morong, W.}, \bibinfo{author}{Oi, D.} \&
  \bibinfo{author}{Ling, A.}
\newblock \bibinfo{title}{{Quantum optics for space platforms}}.
\newblock \emph{\bibinfo{journal}{Opt. and Photon. News}}
  \textbf{\bibinfo{volume}{23}}, \bibinfo{pages}{42--49}
  (\bibinfo{year}{2012}).

\bibitem{Burnham1970}
\bibinfo{author}{Burnham, D.~C.} \& \bibinfo{author}{Weinberg, D.~L.}
\newblock \bibinfo{title}{{Observation of simultaneity in parametric production
  of optical photon pairs}}.
\newblock \emph{\bibinfo{journal}{Phys. Rev. Lett.}}
  \textbf{\bibinfo{volume}{25}}, \bibinfo{pages}{84--87}
  (\bibinfo{year}{1970}).

\bibitem{Trojek2008}
\bibinfo{author}{Trojek, P.} \& \bibinfo{author}{Weinfurter, H.}
\newblock \bibinfo{title}{{Collinear source of polarization-entangled photon
  pairs at nondegenerate wavelengths}}.
\newblock \emph{\bibinfo{journal}{Appl. Phys. Lett.}}
  \textbf{\bibinfo{volume}{92}}, \bibinfo{pages}{91--93}
  (\bibinfo{year}{2008}).

\bibitem{Cheng2015}
\bibinfo{author}{Cheng, C.}, \bibinfo{author}{Chandrasekara, R.},
  \bibinfo{author}{Tan, Y.~C.} \& \bibinfo{author}{Ling, A.}
\newblock \bibinfo{title}{{Space-qualified nanosatellite electronics platform
  for photon pair experiments}}.
\newblock \emph{\bibinfo{journal}{J. Lightwave Technol.}}
  \textbf{\bibinfo{volume}{33}}, \bibinfo{pages}{4799--4804}
  (\bibinfo{year}{2015}).

\bibitem{gomx2}
\bibinfo{author}{NASA}.
\newblock \bibinfo{title}{{NanoRacks-GOMX-2: Small Photon Entangling Quantum
  System (NanoRacks-GOMX-2)}}  (\bibinfo{year}{2015}).
\newblock
  \urlprefix\url{http://www.nasa.gov/mission_pages/station/research/experiments/1328.html#operations}.

\bibitem{NASA2015}
\bibinfo{author}{NASA}.
\newblock \bibinfo{title}{{NASA Independent Review Team Orb-3 Accident
  Investigation Report Executive Summary}}  (\bibinfo{year}{2015}).
\newblock
  \urlprefix\url{http://www.nasa.gov/sites/default/files/atoms/files/orb3_irt_execsumm_0.pdf}.

\bibitem{Tang2014}
\bibinfo{author}{Tang, Z.} \emph{et~al.}
\newblock \bibinfo{title}{{Near-space flight of a correlated photon system.}}
\newblock \emph{\bibinfo{journal}{Sci. Rep.}} \textbf{\bibinfo{volume}{4}},
  \bibinfo{pages}{6366} (\bibinfo{year}{2014}).

\end{thebibliography}

\end{document}